\documentclass{llncs}

\author{David Basin\inst{1} \and Andreas F\"urst\inst{1} \and Thai Son Hoang\inst{1} \and 
  Kunihiko Miyazaki\inst{2} \and Naoto Sato\inst{2}}
\institute{Institute of Information Security, ETH Zurich
  \and Yokohama Research Lab, Hitachi}

\newcommand\PREAMBLE{}
\usepackage[cmex10]{amsmath}
\usepackage{url}

\usepackage{tikz}
\usepackage{tikz-eventB}
\usepackage[color,small]{eventB}
\usepackage{bsymb}
\usepackage{graphicx}
\usepackage[compact]{reqdoc}

\usepackage[plainpages=false]{hyperref}

\hypersetup{
  pdftitle={Abstract Data Types in Event-B -- An Application of Generic
  Instantiation},
  pdfauthor={
    David Basin, Andreas F\"urst, Thai Son Hoang, Kunihiko Miyazaki, Naoto Sato
    Institute of Information Security, ETH Zurich, Switzerland
    Yokohama Research Lab, Hitachi
  },
  pdfsubject={Event-B, Generic Instantiation},
  pdfkeywords={Event-B, Generic Instantiation, Tool, Industrial Case Study},
  colorlinks=true
}

\begin{document}

\title{Abstract Data Types in Event-B -- An Application of Generic Instantiation}
\date{\today}

\maketitle

%!TEX root = adt.tex
\begin{abstract}
  Integrating formal methods into industrial practice is a challenging
  task.  Often, different kinds of expertise are required within the
  same development.  On the one hand, there are domain engineers who
  have specific knowledge of the system under development. On the
  other hand, there are formal methods experts who have experience in
  rigorously specifying and reasoning about formal systems.
  Coordination between these groups is
  important for taking advantage of their expertise.  In this paper,
  we describe our approach of using generic instantiation to
  facilitate this coordination. In particular, generic instantiation
  enables a separation of concerns between the different parties involved
  in developing formal systems.
 \end{abstract}

%%% Local Variables: 
%%% mode: latex
%%% TeX-master: "adt.tex"
%%% End: 

%!TEX root = adt.tex
\section{Introduction}
\label{sec:introduction}

Event-B is a formal method for modelling safe and reliable systems.  Industrial awareness of Event-B has been enhanced by recent collaboration projects (e.g., DEPLOY~\cite{project:_indus}).  These projects acted as a bridge for deploying research results in various industrial contexts with considerable success.  Moreover, they also highlighted several challenges in integrating formal methods into industrial development processes.  In particular, questions about interactions between developers with different kinds of expertise often arise during the deployment.  On the one hand, engineers have domain knowledge including how the systems should work and why they work, but often find it challenging to formalise their reasoning.  On the other hand, formal method experts, which do not have inside knowledge about the specific systems, have experience in reasoning formally about systems in general.

% Contribution
In this paper, we propose adapting the concept of abstract data types to Event-B to enable the interaction between the domain and formal methods experts.  Abstract data types allow developers to hide implementation details that are initially irrelevant to the development of a system.  As a result, systems developed with abstract data types are more intuitive and easier to verify.  The realisation of the abstract data types can be done via generic instantiation by Event-B experts.  In particular, the choice of which (concrete) data structure to use to represent the abstract data type can be done independently of the actual system under development.  Later, generic instantiation in Event-B enables the Event-B expert to prove that the chosen data structure is a valid realisation of the abstract data type.

% Related-work
Generic instantiation in \eventB was introduced in~\cite{DBLP:journals/fuin/AbrialH07} and further elaborated in~\cite{DBLP:conf/icfem/SilvaB09}.  These works show how generic instantiation works with other standard techniques in \eventB such as refinement and composition. This paper illustrates how abstract data types can be modelled and realised using generic instantiation.  Similar to our work is the recently developed \emph{Theory Plug-in}~\cite{maamria:_theor_plug}. The primary usage of the Theory Plug-in is to extend the mathematical language to include new data types.  A theory module also provides an encapsulation of datatypes and enables the separation of concerns between the data types and the models that make use of them.  The main difference between our work and the Theory Plug-in is that data types are usually developed together with their properties within the same theory module.  As a results, the data types developed using the Theory Plug-in are usually already concrete.  There is no clear separation between actual representation of data types and their abstract properties. More information on related work is in Section~\ref{sec:related-work}.

\paragraph{Structure}
In Section~\ref{sec:background} we will give a brief overview of Event-B and generic instantiation in Event-B.  We describe our approach in Section~\ref{sec:contribution}.  In Section~\ref{sec:example} we demonstrate our methodology of splitting the modelling effort on an example. In Section~\ref{sec:related-work} we compare our approach with other existing approaches and in Section~\ref{sec:conclusion} we draw conclusions.

%%% Local Variables: 
%%% mode: latex
%%% TeX-master: "adt.tex"
%%% End: 

%!TEX root = adt.tex
\section{Background}
\label{sec:background}

\subsection{The \eventB Modelling Method}
\label{sec:eventb-modell-meth}

% Event-B general
\eventB \cite{abrial10:_model_in_event_b} is a modelling method for
formalising and developing systems whose components can be modeled as
discrete transition systems.  \eventB is centered around the general
notion of \emph{events} and its semantics is based on transition systems
and simulation between such systems, as described
in~\cite{abrial10:_model_in_event_b}.  We will not describe in detail
the semantics of \eventB here. Instead we just give a brief
description of \eventB models, which are important for generic
instantiation.

% Contexts
\eventB models are organised in terms of two basic constructs:
\emph{contexts} and \emph{machines}. Contexts specify the static part
of a model whereas machines specify the dynamic part. Contexts may
contain \emph{carrier sets}, \emph{constants}, \emph{axioms}, and
\emph{theorems}.  Carrier sets are similar to types. Axioms constrain
carrier sets and constants, whereas theorems are additional properties
derived from axioms.  The role of a context is to isolate the
parameters of a formal model (carrier sets and constants) and their
properties, which are intended to hold for all instances.

% Machine
\emph{Machines} specify behavioural properties of \eventB models.
Machines may contain \emph{variables}, \emph{invariants} (and
\emph{theorems}), and \emph{events}. Variables $v$ define the state of
a machine and are constrained by invariants $I(v)$.  Theorems are
additional properties of $v$ derivable from $I(v)$.  Possible state
changes are described by events.  An event $\Bevt{evt}$ can be
represented by the term \[\eventinline{evt}{}{t}{G(t,v)}{}{S(t,v)}~,\]
where $t$ stands for the event's \emph{parameters}, $G(t,v)$ is the
\emph{guard} (the conjunction of one or more predicates) and $S(t,v)$
is the \emph{action}.  The guard states the necessary condition under
which an event may occur, and the action describes how the state
variables evolve when the event occurs.  We use the short form
$\eventinline{evt}{}{}{G(v)}{}{S(v)}$ when the event does not have any
parameters, and we write $\eventinline{evt}{}{}{}{}{S(v)}$ when, in
addition, the event's guard equals \emph{true}.  A dedicated event
without parameters and guard is used for the \emph{initialisation}
event (usually represented as $\init$).

% Machines see context
A machine can \emph{see} multiple contexts.  During the
development, a context \emph{extends} one or more contexts by declaring
additional carrier sets, constants, axioms or theorems.  An abstract
machine can be \emph{refined} by another concrete machine.  The
variables of the abstract and concrete machines are related by some
gluing invariants.  The existing events are refined accordingly to
this relationship.  Moreover, new events can be added to the concrete
machine.  The new events must refine a special \Bevt{skip}
event, which does not change the abstract variables.

\subsection{Generic Instantiation in \eventB}
\label{sec:gener-inst-eventb}

Generic instantiation is a technique for reusing models by giving
concrete values for abstract parameters of the models.  Generic
instantiation for \eventB is first mentioned
in~\cite{DBLP:journals/fuin/AbrialH07} and is further elaborated
in~\cite{DBLP:conf/icfem/SilvaB09}.  We summarise the approach as
follows.  Suppose we have an abstract development with machines
$\Bmch{M_1}\ldots\Bmch{M_n}$ and their corresponding contexts
$\Bctx{C_1}\ldots\Bctx{C_n}$ as shown in Fig.~\ref{fig:gen-inst-ctx}.
The development is generic, with the carrier sets $s$ and constants
$c$ from the contexts $\Bctx{C_1}\ldots\Bctx{C_n}$ acting as its
parameters.  Assume that $s$ and $c$ are constrained by axioms
$A(s,c)$.
\begin{figure}[!htbp]
  \centering
  \ifx\PREAMBLE\UnDef
\documentclass{beamer}
\usepackage{tikz}
\usepackage[english]{babel}
% or whatever

\usepackage[latin1]{inputenc}
% or whatever
\usepackage{eventB}
\usepackage{tikz-eventB}

\begin{document}
\else
\fi

\begin{tikzpicture}[scale=0.68]
  \footnotesize
  \tikzMch[Mn]{0}{0}{$\Bmch{M_n}(s,c)$}
  \tikzMch[M1]{0}{3}{$\Bmch{M_1}(s,c)$}
  \tikzCtx[Cn]{4.5}{0}{$\Bctx{C_n}(s,c)$}
  \tikzCtx[C1]{4.5}{3}{$\Bctx{C_1}(s,c)$}
  \tikzSees{M1}{C1}
  \tikzSees{Mn}{Cn}
  \tikzRefinesTransitive{Mn}{M1}
  \tikzExtendsTransitive{Cn}{C1}
  \tikzCtx[Dm]{8}{0}{$\Bctx{D_m}(t,d)$}
  \tikzCtx[D1]{8}{3}{$\Bctx{D_1}(t,d)$}
  \tikzExtendsTransitive{Dm}{D1}
  \tikzMch[Nn]{14}{0}{$\Bmch{N_n}(E(t,d),F(t,d))$}
  \tikzMch[N1]{14}{3}{$\Bmch{N_1}(E(t,d),F(t,d))$}
  \tikzRefinesTransitive{Nn}{N1}
  \tikzSees{N1}{Dm}
  \tikzSees{Nn}{Dm}
\end{tikzpicture}

\ifx\PREAMBLE\UnDef
\end{document}
\else
\fi
  \caption{Generic instantiation in \eventB}
  \label{fig:gen-inst-ctx}
\end{figure}
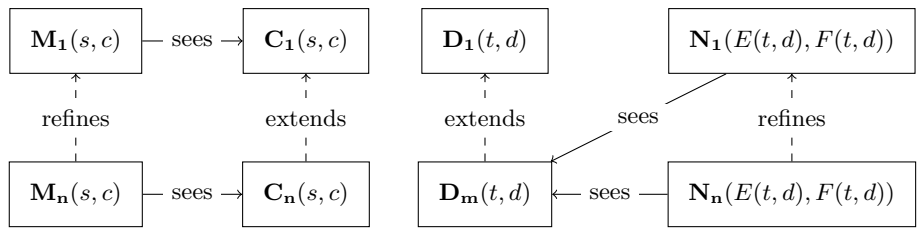

The abstract generic model can be instantiated within another
development containing contexts $\Bctx{D_1}\ldots\Bctx{D_m}$.  Assume
that the concrete contexts $\Bctx{D_1}\ldots\Bctx{D_m}$ contain
concrete carrier sets $t$ and constants $d$, constrained by axioms
$B(t,d)$.  The instantiation is done by giving values for the abstract
carrier sets $s$ and constants $c$ in terms of concrete $t$ and $d$.
Let the concrete expressions $E(t,d)$ and $F(t,d)$ be the instantiated
values for $s$ and $c$ respectively.  Soundness for generic
instantiation requires us to prove that the instantiated abstract
axioms are derivable from the concrete axioms, i.e.,
\[B(t,d) \limp A(E(t,d),F(t,d))~.\] 

In this paper, we further restrict the instantiation for the
abstract carrier sets $s$ so that they can only be instantiated by type-expressions, i.e. $E(t,d)$ must
be some type-expressions.  This is because a
carrier set $S$ in \eventB is assumed to satisfy two additional
constraints (i.e., beside the stated axioms).
\begin{center}
\begin{tabular}{ll}
\textsf{non-empty:} & $S$ is non-empty, i.e., $S \neq \emptyset$.\\
\textsf{maximal:} & $S$ is maximal, i.e. $\forall x \qdot x \in S$.\\
\end{tabular}
\end{center}
The \textbf{maximal} condition is due to the fact that
the \eventB models are typed.  As a result, expressions used for
instantiating carrier sets must be also some type-expressions, i.e.,
satisfying the above two conditions.

Applying generic instantiation, machines \Bmch{N_1}\ldots
\Bmch{N_n} are instances of \Bmch{M_1}\ldots\Bmch{M_n} by
syntactically replacing $s$ and $c$ by $E(t,d)$ and $F(t,d)$.  The
advantage here is that the instantiated machines are
correct by construction.  The resulting model can be used in
conjunction with other techniques such as
refinement~\cite{DBLP:journals/fuin/AbrialH07} and
composition~\cite{DBLP:conf/icfem/SilvaB09}.

%%% Local Variables: 
%%% mode: latex
%%% TeX-master: "adt.tex"
%%% End: 

%!TEX root = adt.tex
\section{Abstract Data Types in Event-B}
\label{sec:contribution}
An abstract data type is a mathematical model of a class of data
structures.  An abstract data type is typically defined in
terms of the operations that may be performed on the data type with
some mathematical constraints on the effects of such operations.  The
advantage of using an abstract data type is that the reasoning can be
done purely based on the properties of the operations, regardless of
the implementation.  We want to use this idea in our developments.
In particular the separation between the abstraction and the
implementation enables us to split the work between domain experts and
formal methods experts.

An abstract data type and its operations can be captured
straightforwardly using contexts in Event-B.  Generic
instantiation can then be used to ``implement'' the abstract data type
and prove that the actual implementation satisfies the constraints on
the effects of the operations.  Our approach can be summarised as follows.
\begin{description}
\item[Domain experts:] The domain experts make use of some
  abstract data types and operations defined within some context to
  model the system in Event-B.  
\item[Formal methods experts:] The formal methods experts use
  generic instantiation to include the details on how the abstract
  data types are represented and prove that the representations
  satisfy the assumptions of the abstract data types stated earlier.
\end{description}

\newBset[ELEMENT]{ELEM}
\newBset[STACKTYPE]{STACK\_TYPE}
\newBcst{STACK}
\newBcst[emptyStack]{empty\_stack}
\newBcst{push}
\newBcst{pop}
\newBaxm[StackType]{axm0\_1}
\newBaxm[emptyStackType]{axm0\_2}
\newBaxm[pushType]{axm0\_3}
\newBaxm[popType]{axm0\_4}
\newBaxm[popDom]{axm0\_5}
\newBaxm[pushNonEmpty]{axm0\_6}
\newBvrb[stack]{s}
\newBvrb[element]{e}
\newBaxm[popPush]{axm0\_7}
\newBvrb[function]{f}
\newBvrb[size]{n}

We illustrate the use of generic instantiation by a model of the
standard \emph{stack data type}. A stack is a last in, first out
(LIFO) data type that contains a collection of elements.  A stack is
characterised by two fundamental operations: \push and \pop.  The
\push operation adds a new item to the top of the stack. The \pop
operation removes the stack's top element.  A special constant
\emptyStack denotes the empty stack.  The stack abstract
data type can be modelled using a context as follows.  Notice that we
have defined the ``type'' \STACKTYPE as a carrier set and the set of
possible stacks \STACK as a constant.
\begin{Bcode}
  $
  \carriersets{\STACKTYPE, \ELEMENT}
  $
  \Bvspace
  $
  \constants{\STACK, \emptyStack, \push, \pop}
  $
  \Bvspace
  $
  \axioms{
    \StackType: & \STACK \subseteq \STACKTYPE \\
    \emptyStackType: & \emptyStack \in \STACK \\
    \pushType: & \push \in \STACK \cprod \ELEMENT \tfun \STACK \\
    \popType: & \pop \in \STACK \pfun \STACK \\
    \popDom: & \dom(\pop) = \STACK \setminus \{\emptyStack\} \\
    \pushNonEmpty: & \forall \stack, \element \qdot \stack \in \STACK
    \quad \limp 
    \quad \push(\stack \mapsto \element) \neq \emptyStack \\
    \popPush: & \forall \stack, \element \qdot \stack \in \STACK
    \quad \limp 
    \quad \pop(\push(\stack \mapsto \element)) = \stack \\
  }
  $
\end{Bcode}

In the representation of stack data type, each stack is represented by
a pair $\function \mapsto \size$, where  $\function$ represents the content of the stack
and $\size$ represents the size of the stack.
Other operations of the stack data type are defined accordingly.  The
concrete context used for instantiation is as follows.  Note that we
use set comprehension to define the constants
accordingly.

\newBaxm[STACKdef]{axm1\_1}%
\newBaxm[emptyStackDef]{axm1\_2}%
\newBaxm[pushDef]{axm1\_3}%
\newBaxm[popDef]{axm1\_4}%
\begin{Bcode}
  $
  \carriersets{\ELEMENT}
  $
  \Bhspace
  $
  \constants{\STACK, \emptyStack, \push, \pop}
  $
  \Bvspace
  $
  \axioms{
    \STACKdef: & \STACK = \{ \function \mapsto \size \mid \size \in
    \nat \land \function \in 1 \upto \size \tfun \ELEMENT \}  \\
    \emptyStackDef: & \emptyStack = \emptyset \mapsto 0 \\
    \pushDef: & \push = \{\function,\size, \element \qdot \\
    & \phantom{\push = \{} \quad \function
    \mapsto \size \in STACK \land \element \in \ELEMENT \quad \mid \\
    & \phantom{\push = \{} \quad ((\function \mapsto \size) \mapsto \element) \mapsto ((\function
    \ovl \{(\size+1) \mapsto \element\}) \mapsto \size+1)\} \\
    \popDef: & \pop = \{\function,\size \qdot \function
    \mapsto \size \in STACK \land \size \neq 0 \quad \mid \\
    & \phantom{\pop = \{} \quad (\function \mapsto \size) \mapsto
    ((\{\size\} \domsub \function) \mapsto \size-1)\}
  }
  $
\end{Bcode}

To prove that the representation of the stack data type is
consistent with the stack abstract data type, we can use instantiation
where the abstract constants are instantiated with concrete
constants with the same name.  The abstract carrier set $\STACKTYPE$
is instantiated with $\pow(\intg \cprod \ELEMENT) \cprod \intg$.  The
abstract axioms (i.e., $\StackType$ -- $\popPush$) must be
derived from the concrete axioms (i.e., $\STACKdef$ --
$\popDef$).  This can be done by expanding the definitions of the
concrete constants accordingly.

%%% Local Variables: 
%%% mode: latex
%%% TeX-master: "adt.tex"
%%% End: 

%!TEX root = adt.tex
\section{Example}
\label{sec:example}
We illustrate our approach by modelling a set of trains on a railway
network, inspired by the example in~\cite[Chapter
17]{abrial10:_model_in_event_b}.

\subsection{Requirements Document}
\label{sec:requ-docum}
\begin{figure}[htbp]
  \begin{center}
    \includegraphics[width=0.9\textwidth]{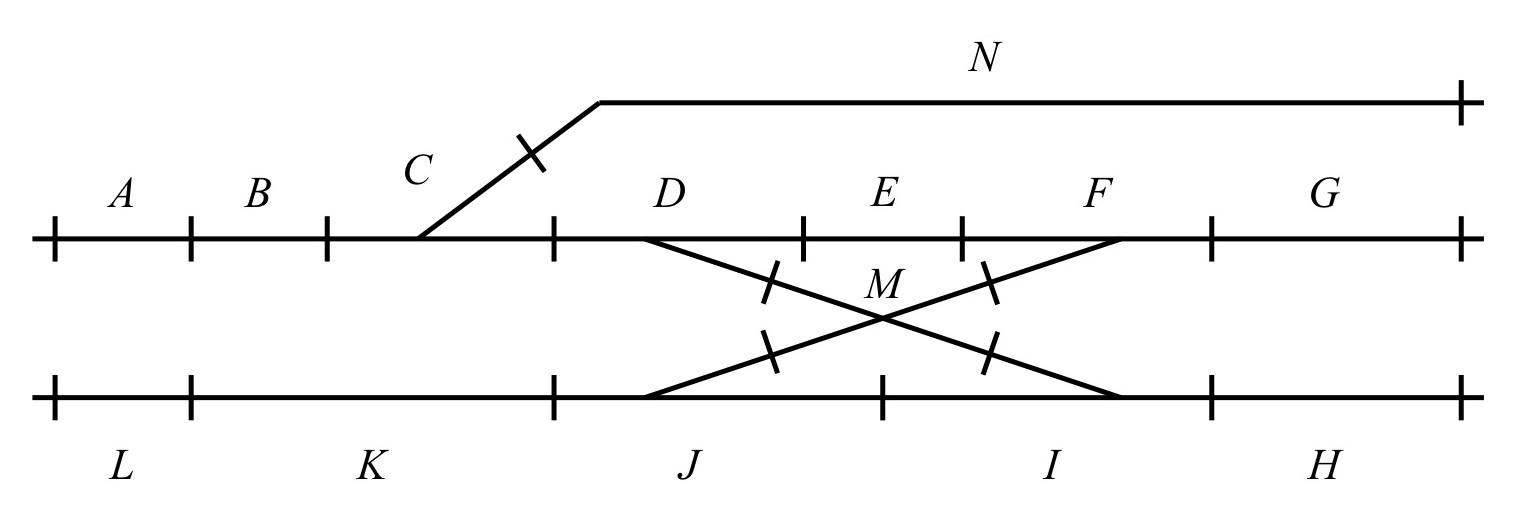}
    \caption{Layout of a sample network with sections A to N.}
    \label{fig:network}
  \end{center}
\end{figure}

A railway network is divided into sections. An example of such a network is showed in Figure~\ref{fig:network}, taken from~\cite[Chapter 17]{abrial10:_model_in_event_b}.

A set of trains are moving within the network.  Two important requirements are that trains must not derail or collide. To avoid collision, the system must ensure that each section is occupied by at most one train. Moreover, trains are assumed to move only forward within the network.

\begin{requirements}
  \saf[0.8\textwidth]{saf:no-collision}{For each section, at most one train occupies that section.}
  \ReqSpacing[1ex]
  \saf{saf:no-derailment}{Trains are always on the network.}
  \ReqSpacing[1ex]
  \asm[0.8\textwidth]{asm:move-forward}{Trains only move forward.}
\end{requirements}

\subsection{Informal Discussion}
\label{sec:some-inform-disc}
An important part of the model will formalise the trains moving
within the network.  Intuitively, a train can be seen as the sequence of
consecutive sections that it occupies within the network.  There are
different possible formalisation of the trains, e.g., using functions
relating occupied sections as in~\cite[Chapter
17]{abrial10:_model_in_event_b}, or modeling sequences as functions
from integers to sections.  However, the system should be correct
regardless of which modelling style is used to represent the
trains.  In particular, the formalisation of the trains in Event-B is
of little interest to the domain experts.  It would be easier for
the domain experts to model the trains at the more abstract level, 
i.e. with a train abstract data type.  The
decision of which representation for the train data type will be
decided by the Event-B experts.  In particular, different
representations can be used for the train data type via separate
instantiation.

% If we asked a domain expert how to
% describe a train the answer could be as follows. In the scope of an
% interlocking system, a train is one single unit that has a front and a
% rear. It occupies some sections of the underlying track network. It
% enters the network in one section and can move forward. This
% description is very abstract and does not provide any details on what
% data structure should be used for representing a train.

% An Event-B expert knows what data structure might best fit such a
% model so that as many proofs as possible are automatically discharged
% by the Rodin platform's built-in theorem provers. A field expert,
% however, developing a system should not have to bother with the
% peculiarities of Event-B and the Rodin platform. This task of finding
% a good data structure should therefore be assigned to an Event-B
% expert, while the field expert can model the desired system at an
% abstract level. In the following we present how a field expert could
% describe a train and model it in Event-B at such an abstract level.

\newBset[TRAINTYPE]{TRAIN\_TYPE}
\newBset{SECTION}
\newBset[TRAINID]{TRAIN\_ID}

\newBcst{TRAIN}
\newBcst{area}
\newBcst{head}
\newBcst{rear}
\newBcst[addHead]{add\_head}
\newBcst{front}
\newBcst[newTrain]{new\_train}
\newBvrb{trains}
\newBevt[extendHead]{extend\_head}
\newBevt{enter}
\newBevt[removeRear]{remove\_rear}
\newBcst{connection}
\newBpar[train]{t}
\newBpar[sect]{s}
\newBaxm[areaType]{area\_Type}
\newBaxm[headType]{head\_Type}
\newBaxm[rearType]{rear\_Type}
\newBaxm[addHeadType]{add\_head\_Type}
\newBaxm[frontType]{front\_Type}
\newBaxm[newTrainType]{new\_train\_Type}
\newBaxm[areaAddHead]{area\_add\_head}
\newBaxm[areaFront]{area\_front}
\newBaxm[areaNewTrain]{area\_new\_train}
\newBaxm[trainDef]{train\_Def}
\newBaxm[headDef]{head\_Def}
\newBaxm[rearDef]{rear\_Def}
\newBaxm[areaDef]{area\_Def}
\newBaxm[addHeadDef]{add\_head\_Def}
\newBaxm[frontDef]{front\_Def}
\newBaxm[newTrainDef]{new\_train\_Def}
\newBaxm[connectionDef]{connection\_Def}

\newBaxm[addHeadDom]{add\_head\_dom}
\newBaxm[frontDom]{front\_dom}
\newBaxm[connectionType]{connection\_Type}
\newBaxm[connectionAddHead]{connection\_add\_head}
\newBaxm[connectionFront]{connection\_front}
\newBaxm[connectionNewTrain]{connection\_new\_train}
\newBcst{NETWORK}

\newBinv[collisionFree]{collision\_free}

\subsection{Formal Model}
\label{sec:form-model}

\subsubsection{Train Abstract Data Type}
\label{sec:train-abstract-data}
\begin{figure}[htbp]
  \begin{center}
    \includegraphics[width=0.9\textwidth]{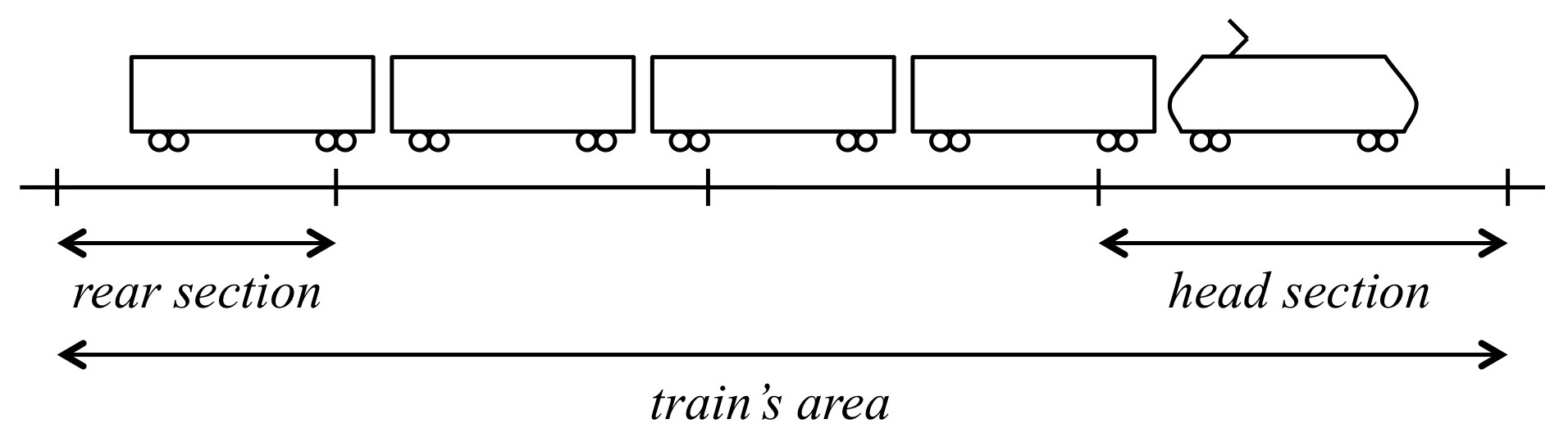}
    \caption{Train in the network occupying sections.}
    \label{fig:train}
  \end{center}
\end{figure}

We first formalise the train abstract data type in a context,
focusing on requirement~\ref{saf:no-collision}.  In particular,
we consider the following ``attributes'' of a train: the sections that
the train occupies (we refer to them as the train's area), the section
of the train's head (the end where the train driver is sitting) and
the section of the train's rear (the opposite end). This is
illustrated in Figure~\ref{fig:train}.

Let the set of sections be a carrier set \SECTION.  We abstractly
represent the trains state by a constant \TRAIN, that is a subset of the
carrier set \TRAINTYPE.  Three function constants, namely \area, \head, and
\rear, are used to get the information about the trains' area, head
position, and rear position, respectively.  For an abstract data type
describing a train, one can see these constants as operations of the
data type.
\begin{center}
\begin{tabular}{ll}
\area : & takes a train state, returns a set of sections.\\
\head : & takes a train state, returns a section.\\
\rear : & takes a train state, returns a section.\\
\end{tabular}
\end{center}
In \eventB, we give the typing information for these constants using
the following axioms.
\begin{Bcode}
  $
  \axioms[off]{
    \areaType: & \area \in \TRAIN\tfun\pow (\SECTION) \\
    \headType: & \head \in \TRAIN\tfun \SECTION \\
    \rearType: & \rear \in \TRAIN\tfun \SECTION
  }
  $
\end{Bcode}
Further constraints on these constants will be given later when they are needed for
maintaining the correctness of the machines that use this data
type.

%  In the following
% we will refer to these kinds of functions as \emph{observer
%   functions}. We need this functions as we do not know yet how to
% access the state of a system as we postpone the decision of what data
% structure to use.
% \begin{align*}
% area &\in TRAIN\tfun\pow (SECTION) \\
% head &\in TRAIN\tfun SECTION \\
% rear &\in TRAIN\tfun SECTION \\
% \end{align*}

% The definition of the observer functions so far specifies only their
% types. It is not defined yet how the sections are retrieved from a
% given train. This will be defined later by axioms.

When a train moves, the set of sections it occupies changes. When
moving forward,~\ref{asm:move-forward}, the train's head reaches the
end of its head section and moves to the new section ahead. Similarly,
when the train's rear leaves the train's rear section, the rear is
reassigned. The train's area is updated accordingly: it is extended to
include the new head section when the head moves, and the
rear section is removed when the rear moves. As a result, we define two
additional operations for manipulating the train.
\begin{center}
\begin{tabular}{ll}
\addHead : & takes a train state and a section, returns a
  train state.\\
\front : & takes a train state, returns a train state.\\
\end{tabular}
\end{center}
In \eventB, we give the type for these constant as follows.
\begin{Bcode}
  $
  \axioms[off]{
     \addHeadType: & \addHead \in \TRAIN\cprod \SECTION\pfun \TRAIN \\
     \frontType: & \front \in \TRAIN \pfun \TRAIN
  }
  $
\end{Bcode}
Note that we use partial functions to indicate that there are
some constraints for extending the train's head and removing the train's
rear.

% e.g., the train's head
% eventually reaches the border of its head section. When moving
% further, and therefore crossing the border, the head is reassigned to
% the new section ahead. Similarly, as soon as the train's rear leaves a
% section, this section is removed from the train's area and the rear is
% reassigned.

% In our abstraction we do not specify the updating of a train's head
% and rear location. This would be too specific and would favour some
% data structures. Instead, we exchange the train with a new one from
% the set of all possible trains. At the abstract level, we are not
% interested in what way the train is updated. We only care what a given
% train looks like after some operation (as moving forward). For this,
% we define \emph{ functions} that return a new train for a
% given one. In contrast to the observer functions, the manipulation
% function are in general only partial functions. The manipulation
% function for moving forward to a new head section \emph{add\_head} is
% defined as follows.

Finally, we define an additional operation \newTrain to create a new
train when the train enters the network from a particular section.
\begin{Bcode}
  $
    \axioms[off]{
       \newTrainType: & \newTrain\in \SECTION\tfun \TRAIN
    }
  $
\end{Bcode}

\subsubsection{System Model Using Train Abstract Data Type}
\label{sec:system-model-using}
Using the train abstract data type, the system can be straightforwardly
modelled.  Let \TRAINID be the set of possible IDs for trains in
the network. The variable \trains represents the trains currently
monitored by the systems, which is a mapping from train IDs to actual
trains.  Initially, \trains is assigned the empty set $\emptyset$.
\begin{Bcode}
  $
  \variables{\trains}
  $
  \Bhspace
  $
  \invariants{
    \trains \in \TRAINID \pfun \TRAIN
  }
  $
\end{Bcode}
Three events \enter, \extendHead, \removeRear are used to model
the different cases where a train enter the network, a train extends
its head to a new section, and a train removes its rear section.
\begin{Bcode}
  $
  \event{\enter}{}{\train, \sect}{\train \notin \dom(\trains) \\
    \sect \in \SECTION}{}{\trains(\train) \bcmeq \newTrain(\sect)}
  $
  \Bhspace
  $
  \event{\extendHead}{}{\train, \sect}{
    \train \in \dom(\trains) \\
    \sect \notin \area(\trains(\train))
  }{}{
    \trains(\train) \bcmeq \addHead(\trains(\train) \mapsto \sect)
  }
  $
  \Bvspace
  $
  \event{\removeRear}{}{\train}{
    \train \in \dom(\trains) \\
    \head(\trains(\train)) \neq \rear(\trains(\train))
  }{}{
    \trains(\train) \bcmeq \front(\trains(\train))
  }
  $
\end{Bcode}
In particular the guard of \extendHead states that the new section
\sect is not already occupied by the train \train, and the guard of
\removeRear states that the head and the rear of the train $\train$
are in different sections.  Moreover, these events lead us to the
following constraints about the domain of operations \addHead and
\front.
\begin{Bcode}
  $
  \axioms[off]{
    \addHeadDom: & \dom(\addHead) = \{\train \mapsto \sect \mid \train
    \in \TRAIN \land \sect \notin \area(\train)\} \\
    \frontDom: & \dom(\front) = \{\train \mid \train \in
    \TRAIN \land \head(\train) \neq \rear(\train)\}
  }
  $
\end{Bcode}

An important invariant captures requirement~\ref{saf:no-collision},
stating that for any two distinct trains $\train_1$, $\train_2$, they do not
occupy the same section.
\begin{Bcode}
  $
  \invariants[off]{
    \forall \train_1, \train_2 \qdot \train_1 \in \dom(\trains) \land
    \train_2 \in \dom(\trains) \land \train_1 \neq \train_2 ~\limp~ \\
    \quad \area(\trains(\train_1)) \binter \area(\trains(\train_2)) = \emptyset
  }
  $
\end{Bcode}

The invariant leads to the following additional guard for \enter{} and \extendHead
\[\forall \train_1 \qdot \train_1 \in \dom(\trains) ~\limp~ \sect
\notin \area(\trains(\train_1))\]

While proving the correctness of our model, we discovered the following required constraints on the
train abstract data type. These constraints are formalised by additional axioms over the abstract data type's
operations.

\begin{Bcode}
  $
  \axioms[off]{
    \areaAddHead: & \forall \train, \sect \qdot \train \mapsto \sect \in
    \dom(\addHead) ~\limp \\
    & \hspace{10em} \area(\addHead(\train \mapsto \sect)) =
    \area(\train) \bunion \{\sect\} \\
    \areaFront: & \forall \train \qdot \train \in \dom(\front) ~\limp~
    \area(\front(\train)) = \area(\train) \setminus \{\rear(\train)\}
    \\
    \areaNewTrain: & \forall \sect \qdot \sect \in \SECTION ~\limp~
    \area(\newTrain(\sect)) = \{\sect\}
  }
  $
\end{Bcode}

In order to specify the fact that the trains do not
derail,~\ref{saf:no-derailment}, we introduce another operation,
\connection, on the train abstract data type to specify the connections
of the sections belonging to a train.  The typing information for
\connection is as follows.
\begin{Bcode}
  $
  \axioms[off]{
    \connectionType: & \connection \in \TRAIN \tfun (\SECTION
    \rel \SECTION)
  }
  $
\end{Bcode}
The invariant corresponding to ~\ref{saf:no-derailment} is
\begin{Bcode}
  $
  \invariants[off]{
    \forall \train \qdot \train \in \dom(\trains) ~\limp~
    \connection(trains(\train)) \subseteq \NETWORK
  }
  $,
\end{Bcode}
where $\NETWORK$ is a constant describing the topology of the actual
network.  An additional guard is added to event \extendHead as
follows.
\[\sect \mapsto \head(\trains(\train)) \in \NETWORK\]

Again, we discovered additional constraints on the operation \connection
while proving the model.
\begin{Bcode}
  $
  \axioms[off]{
    \connectionAddHead: & \forall \train, \sect \qdot \train \mapsto \sect \in \dom(\addHead)
    ~\limp~ \\
    & \quad \connection(\addHead(\train \mapsto \sect)) =
    \connection(\train) \bunion \{\sect \mapsto \head(\train)\} \\
    \connectionFront: & \forall \train \qdot \train \in \dom(\front) ~\limp~
    \connection(\front(\train)) \subseteq \connection(\train) \\
    \connectionNewTrain: & \forall \sect \qdot \sect \in \SECTION ~\limp~
    \connection(\newTrain(\sect)) = \emptyset
  }
  $
\end{Bcode}
Note that axiom \connectionFront does not specify exactly how a train's
connection is changed when the rear is removed. It only
specifies that the connection will not be enlarged.  This suffices
for proving the no-derailment property of the system.

% As before, this is only typing information and does not describe what
% the new train looks like when adding a new head section.  The
% differences of the new train in contrast to the old one are as
% follows.
% \begin{itemize}
% \item The rear of the train remains in the same section.
% \item The head of the train is pointing to the new section.
% \item The new section is added to the train's area.
% \end{itemize}

\subsubsection{Generic Instantiation}
\label{sec:gener-inst}
We now need to find a representation for the train data
type.  This is the point where the role of the formal method expert
becomes prominent.  As mentioned before, different data structures can
be used to represent the train abstract data type.  We present here a
solution where a train is represented by a function from an integer interval to the
set of sections.  Each train is associated with a tuple $(a,b,f)$,
where the interval $a \upto b$ represents the domain of a total injective
function $f$.
\begin{Bcode}
  $
    \axioms[off]{
       \trainDef: & \TRAIN = \{ a\mapsto b\mapsto f\mid a\in\intg\land a\leq b\land f\in a\upto b\tinj \SECTION \}
    }
  $
\end{Bcode}

The train's head is located at the lower end of the interval ($a$) and
its rear at the upper end ($b$). Injectivity guarantees that the
sequence cannot include a section twice at different positions.  The
operations on the train data type are defined accordingly.
\begin{Bcode}
  $
    \axioms[off]{
       \headDef: & \head = \{a,b,f\cdot a\mapsto b\mapsto f\in \TRAIN\mid (a\mapsto b\mapsto f)\mapsto f(a)\}\\
       \rearDef: & \rear = \{a,b,f\cdot a\mapsto b\mapsto f\in \TRAIN\mid (a\mapsto b\mapsto f)\mapsto f(b)\}\\
       \areaDef: & \area = \{a,b,f\cdot a\mapsto b\mapsto f\in
       \TRAIN\mid (a\mapsto b\mapsto f)\mapsto f[a\upto b]\}\\
       \addHeadDef: & \addHead = \{ a,b,f,s\cdot a\mapsto b\mapsto f\in \TRAIN\land s\notin t[a\upto b] \\
       & \hspace{2em} \mid (a\mapsto b\mapsto f)\mapsto s\mapsto ((a-1)\mapsto b\mapsto (f\bunion\{ a-1\mapsto s\}))\}\\
       \frontDef: & \front = \{ a,b,f\cdot a\mapsto b\mapsto f\in \TRAIN\land a\neq b \\
       & \hspace{2em} \mid (a\mapsto b\mapsto f)\mapsto (a\mapsto (b-1)\mapsto (\{ b\}\domsub f))\}\\
       \newTrainDef: & \newTrain = \{ s\cdot s\in \SECTION\mid s\mapsto(1\mapsto 1\mapsto \{1\mapsto s\})\}\\
       \connectionDef: & \connection = \{a,b,f\cdot a\mapsto b\mapsto f\in \TRAIN \\
      & \hspace{2em} \mid (a\mapsto b\mapsto f)\mapsto\{i\cdot i\in a\upto b-1\mid t(i)\mapsto t(i+1)\}\} \\
    }
  $
\end{Bcode}

By instantiating the abstract type \TRAINTYPE to $\intg \cprod \intg
\cprod \pow(\intg \cprod \SECTION)$ and other abstract constants with
the concrete constants of the same name, we can prove that the
constraints of the train abstract data type (abstract axioms) are
derivable from the definition of the train data type.

For instantiating the train abstract data type, we used the prototype plug-in for generic instantiation.

\section{Related Work}
\label{sec:related-work}
Generic instantiation in \eventB has been introduced
in~\cite{DBLP:journals/fuin/AbrialH07} and is further elaborated
in~\cite{DBLP:conf/icfem/SilvaB09}.  Both papers illustrate the use of
generic instantiation for reusing formal models by combining it with
existing techniques like refinement and composition.  In this paper, we
illustrate another application of generic instantiation for
algebraically modelling abstract data types.  In particular, the
abstract development and the concrete instantiated development enable
the separation of concerns between domain experts
and formal methods experts.  The domain experts can work with the
abstract models, stating the assumptions under which the systems 
work correctly.  The formal method experts use 
generic instantiation to prove that the actual implementations satisfy the
assumptions as required by the domain experts.

A similar form of generic instantiation is also available in classical
B\cite{abrial96:_b}.  A development in classical B
also contains abstract data which must be finalised when
the final software products are deployed.  This finalisation process
is an instantiation step, involving validating that the actual data
satisfies the assumptions stated in the formal
model~\cite{DBLP:journals/fac/LeuschelFFP11}.  We illustrate here
(together with other
work~\cite{DBLP:journals/fuin/AbrialH07,DBLP:conf/icfem/SilvaB09}) that
generic instantiation is also useful during the stepwise development of the
formal models, not just as the last realisation step in deploying the
formal models.

Recent development of the \emph{Theory
  Plug-in}~\cite{maamria:_theor_plug} allows users to extend the
mathematical languages of \eventB, e.g., by including new data types.
Theorems about new data types can be stated and used later by a
dedicated tactic associated with the Theory Plug-in.  There is also a
clear distinction between the theory modules (capturing data
structures and their properties) and the \eventB models making use of
the newly defined data structures.  This distinction also enables a
collaboration between domain experts and
formal methods experts: the domain experts work with the \eventB
models while the formal methods experts work with the theory modules.
The difference with our approach is the order in which the work is
carried out.  With the Theory Plug-in, the domain experts rely on the
theory developed by the formal methods experts.  In our approach, the
input for the formal methods experts are the abstract models that are
developed by the domain experts, including the assumptions stated as
axioms on the abstract carrier sets and constants.  Another difference
is that we can have different implementations for the abstract data
types.

Our approach is similar to work on algebraic
specification~\cite{DBLP:journals/fac/SannellaT97}.  In this domain, a
specification contains a collection of \emph{sorts}, \emph{operations}, 
and \emph{axioms} constraining the operations.  Specifications
can be \emph{enriched} by additional sorts, operations, or axioms.
Furthermore, to develop programs from specifications, the
specifications are transformed via a sequence of small refinement
steps.  During these steps, the operations are ``coded'' until the
specification becomes a concrete description of a program.  For each such
refinement step, it is required to prove that the code of the
operations satisfy the axioms constraining them.  An algebraic
specification therefore corresponds to an \eventB context, while the
refinement of the algebraic specifications is similar to generic
instantiation in \eventB.  The main difference between algebraic
specification and \eventB is that there is no corresponding elements
to \eventB machines.  In particular, we make use of the dynamic
information of \eventB machines to derive the necessary axioms on the
abstract data types.

%%% Local Variables: 
%%% mode: latex
%%% TeX-master: "adt.tex"
%%% End: 

%!TEX root = adt.tex
\section{Conclusion and Future Work}
\label{sec:conclusion}
In this paper we presented our approach to modeling abstract data types and their implementation in Event-B.  Using abstract data types allows us to hide irrelevant details that are not important for the domain expert. The domain expert can focus on modelling the functionality of the system which is his core competence.  Abstract data types thereby have a similar purpose to programming interfaces in programming languages.  The instantiation of the abstract data type is left to an Event-B expert.  The way we introduced the concept of abstract data types in our approach allows us to utilise generic instantiation which handles both the substitution of the abstract data type by the chosen data structure as well as the generation of the needed proof obligations to guarantee that the chosen structure is a valid instance of the abstract data type.

We successfully applied our approach to the example in this paper as well as a substantially more complex version of it. Further investigation is needed on the scalability of the approach, which is essential for its applicability in industrial development processes. Furthermore, we are interested in applying our approach outside the domain of railway systems to obtain evidence for its generality.

%%% Local Variables: 
%%% mode: latex
%%% TeX-master: "adt.tex"
%%% End: 

\bibliography{adt}

\end{document}